\documentclass[aps,pre,showpacs]{revtex4-2}
\usepackage{color}
\usepackage{amssymb}
\usepackage{amsmath}
\usepackage{graphicx}
\usepackage{bm}

\begin{document}
\title{Transient inverse energy cascade in free surface turbulence}

\author{G. Boffetta}
\affiliation{Dipartimento di Fisica and INFN, Universit\`a degli Studi di Torino, via P. Giuria 1, 10125 Torino, Italy.}

\author{A. Mazzino}
\affiliation{Dipartimento di Ingegneria Civile, Chimica e Ambientale and INFN, Universit\`a degli Studi di Genova, via Montallegro 1, 16145 Genova, Italy.}

\author{S. Musacchio}
\affiliation{Dipartimento di Fisica and INFN, Universit\`a degli Studi di Torino, via P. Giuria 1, 10125 Torino, Italy.}

\author{M. E. Rosti}
\affiliation{Complex Fluids and Flows Unit, Okinawa Institute of Science and Technology Graduate University, 1919-1 Tancha, Onna-son, Okinawa 904-0495, Japan}

\begin{abstract}
We study the statistics of free-surface turbulence at large Reynolds numbers
produced by direct numerical simulations in a fluid layer 
at different thickness
with fixed characteristic forcing scale.
We observe the production of a transient inverse cascade, with a duration
which depends on the thickness of the layer, followed by a transition to three-dimensional
turbulence initially produced close to the bottom, no-slip boundary. 
By switching off the forcing, we study the decaying turbulent regime 
and we find that it cannot be described by an exponential law. 
Our results show that boundary conditions play a fundamental role in the
nature of turbulence produced in thin layers and give limits on the 
conditions to produce a two-dimensional phenomenology.
\end{abstract}


\maketitle

\section{Introduction}
\label{sec1}

Many geophysical and astrophysical flows are confined in thin
layers of small aspect ratio either by material boundaries or by 
physical mechanisms which constrain the motion, such as rotation or
stratification. 
The vertical extension (thickness) of such layers can be much 
smaller than the typical horizontal scales, while being at the same 
time much larger than the dissipative viscous scales. As a consequence,
turbulent flows in those quasi-two-dimensional geometries display a 
rich phenomenology with both three-dimensional (3D) features at small scales
and two-dimensional (2D) properties at large scales.

Remarkably, numerical simulations have shown that a physical 
confinement is not necessary to observe a two-dimensional phenomenology.
Even fully periodic simulations in a box with large aspect ratio 
$L_z \ll L_x=L_y$, forced at intermediate scales, 
produce a split energy cascade in which a
fraction of the energy flow to large scales (as in a pure 2D flow) and the
remaining part goes to small scales producing the 3D direct cascade
\cite{smith1996crossover,celani2010turbulence,alexakis2018cascades}.
In this case, the key parameter which controls the relative flux of 
energy in the two cascades is the ratio $S=L_z/L_f$ between the confining
scale and the characteristic scale of the forcing $L_f$.
In the limit $S \to 0$ of very small thickness, vertical motion is
suppressed by viscosity and the flow fully recovers the 2D 
phenomenology
\cite{musacchio2017split}.
The phenomenology of this energy split can be changed by other physical 
ingredients, including solid body rotation \cite{deusebio2014dimensional}
and stratification \cite{sozza2015dimensional}. Moreover, the 
effects of the transition are important for several statistics, including
the chaoticity of the flow \cite{clark2021chaotic}.
Since at larger scale the flow becomes more two-dimensional, 
the inverse cascade of energy in thin layer proceeds until it reaches 
the largest scale and produces a large-scale structure, called the 
condensate \cite{xia2009spectrally,laurie2014universal,musacchio2019condensate,van2019condensates,fang2021spectral}.

The phenomenology changes in the presence of confining physical 
boundaries, as in the case of laboratory experiments performed in a
thin layer of fluid confined by gravity
\cite{shats2010turbulence,byrne2011robust,xia2011upscale}.
The bottom (and lateral) boundary of the tank produces a boundary
layer which dissipates a relevant fraction of the energy injected in the system
\cite{van2009studies,shats2010turbulence} and thus reduces the turbulent 
flux, in particular in the case of a single layer of 
fluid \cite{byrne2011robust}. 
Experiments with a double layer, in particular of immiscible fluids,
reduce the damping rate induced by the bottom wall, and produce an inverse 
cascade of energy \cite{chen2006physical,byrne2011robust}.

In this work we systematically study the turbulent flow in a 
thin layer by extensive direct numerical simulations 
of the Navier-Stokes equations in a box with no-slip 
boundary conditions (BC) at the bottom, and a free-slip
BC at the top, similarly to laboratory experiments of 
free-surface turbulence. 
Simulations are done at high resolution
and large Reynolds numbers which allows the development of a 
fully 3D turbulent motion. The forcing scale $L_f$ is fixed, and 
we vary the ratio $S$ by changing the thickness $L_z$ of the box.
We find that, in the range of parameters explored here, the thin
layer is unable to sustain an inverse cascade of energy.
We observe a transient inverse cascade, of duration which depends
on the thickness $S$, but the 3D motion in the bottom boundary layer 
eventually propagates to the full layer and the flow becomes fully 
three-dimensional with the inverse cascade being suppressed. 
We also study the decaying regime of our systems, and we 
find a complex behavior which cannot be described by a simple
exponential law.

The remaining of this paper is organized as follows. In Sec.~\ref{sec2},
we present the details of the numerical simulations. Section~\ref{sec3}
is devoted to the evolution of the global quantities of the flow, 
while Sec.~\ref{sec4} discusses small scale statistics and in particular
the presence of a direct or inverse cascade. In Sec.~\ref{sec5}, we report
the results of the decaying simulations and finally Sec.~\ref{sec6} is
devoted to the conclusions.

\section{Models and direct numerical simulations}
\label{sec2}

We consider the 3D Navier-Stokes equations for an incompressible
velocity field ${\bm u}({\bm x},t)=(u,v,w)$
(with ${\bm \nabla} \cdot {\bm u}=0$) in a domain of dimension
$L_x \times L_y \times L_z$
\begin{equation}
{\partial {\bm u} \over \partial t} + {\bm u} \cdot {\bm \nabla} {\bm u} =
- {\bm \nabla} p + \nu \nabla^2 {\bm u} + {\bm f},
\label{eq1}
\end{equation}
where the constant density has been adsorbed into the pressure $p$ and
$\nu$ is the kinematic viscosity.
The two-dimensional forcing ${\bm f}$ is restricted to the two
horizontal ($x,y$) components (2D2C) ${\bm f}({\bm x})=(f_x(x,y),f_y(x,y),0)$.
It is Gaussian, white in time, and in Fourier space is confined
in a narrow cylindrical shell of wavenumbers centered around
$k_f=2 \pi/L_f=8$. One reason to have a $2D$ forcing is that it is 
independent on the thickness of the flow and therefore we use
the same forcing for all the simulations at different $L_z$. 
Moreover, thanks to the delta-correlation in time, the rates of injection of
energy $\varepsilon$ is fixed and does not depend on $L_z$ or 
on the properties of the flow.

Boundary conditions are periodic in the horizontal direction $(x,y)$
while, to simulate free-surface turbulence, we impose a no-slip BC
at the bottom $z=0$ and a free-slip BC at the top
$z=L_z$. We have therefore $u=v=w=0$ 
at $z=0$
while $\partial_z u=\partial_z v=0$ and $w=0$ at $z=L_z$. 
We remark that these BC have been previously used for 
numerical studies of free-surface turbulence \cite{nagaosa1999direct}.

The equations of motion are solved numerically with a uniform spacing in all
directions. To solve the problem, we use the flow solver \textit{Fujin}, an
in-house code, extensively validated and used in a variety of
problems~\cite{olivieri2020dispersed, rosti2021shear, mazzino2021unraveling,
brizzolara2021fiber, mazzino2022puff, hori2022eulerian}, based on the
(second-order) finite-difference method for the spatial discretization and the
(second-order) Adams-Bashforth scheme for time marching. See also
\texttt{https://groups.oist.jp/cffu/code} for a list of validations.
Simulations are performed at a fixed horizontal resolution and varying
the vertical resolution depending on $S$ with a constant viscosity and 
energy input (see Table~\ref{table1}). 
Additional simulations at different viscosities (not
discussed here) produced similar results. 
All the simulations start from an initial zero-velocity field 
and reach a statistically stationary states characterized 
by a constant energy. Another set of simulations starts from 
these asymptotic states and study the decaying regime by switching
off the forcing. 
We remark that since $\langle f_i \rangle=0$
(the average is defined over the $(x,y)$ planes) we have no mean flow
$\langle u_i \rangle=0$ and we will consider the statistics of
the fluctuating velocity field only.

\begin{table}[h!]
\begin{tabular}{cccccc}
$Run \#$ & $N_z$ & $S$ & $E_{2D}$ & $E_z$ & $Re$ \\ \hline
$1$ & $16$ & $0.03125$ & $3.9$ & $0.001$ & $1580$ \\
$2$ & $32$ & $0.0625$ & $7.8$ & $0.07$ & $2250$ \\
$3$ & $48$ & $0.09375$ & $9.9$ & $0.22$ & $2550$ \\
$4$ & $64$ & $0.125$ & $11.6$ & $0.41$ & $2780$ \\
$5$ & $96$ & $0.1875$ & $12.6$ & $0.68$ & $2920$ \\
$6$ & $128$ & $0.25$ & $13.9$ & $0.96$ & $3100$ \\
$7$ & $160$ & $0.3125$ & $12.4$ & $1.1$ & $2940$ \\
$8$ & $512$ & $1.0$ & $12.5$ & $2.1$ & $3060$  \\
$9$ & $1024$ & $2.0$ & $12.5$ & $2.7$ & $3100$ 
\end{tabular}
\caption{Parameters of the simulations. 
$N_z$ is the resolution in the $z$ direction, $S=L_z/L_f$ the thickness in the $z$ 
direction (in units of the forcing scale $L_f$), $E_{2D}$ and $E_z$ are the 
energies of the horizontal and vertical component of the velocity in the
stationary state (the total energy is $E=E_{2D}+E_z$). The Reynolds
number is defined as $Re=\sqrt{E} L_f/\nu$.
For all the runs $L_x=L_y=2 \pi$ with
a resolution $N_x=N_y=4096$. The viscosity is $\nu=9.8 \times 10^{-4}$ 
and the random forcing is active on the scale $L_f=L_x/8$ with a fixed
energy input $\varepsilon_{I}=50$. 
}
\label{table1}
\end{table}

\section{Large-scale properties of the flow}
\label{sec3}

\begin{figure}[h!]
\centerline{\includegraphics[scale=0.75]{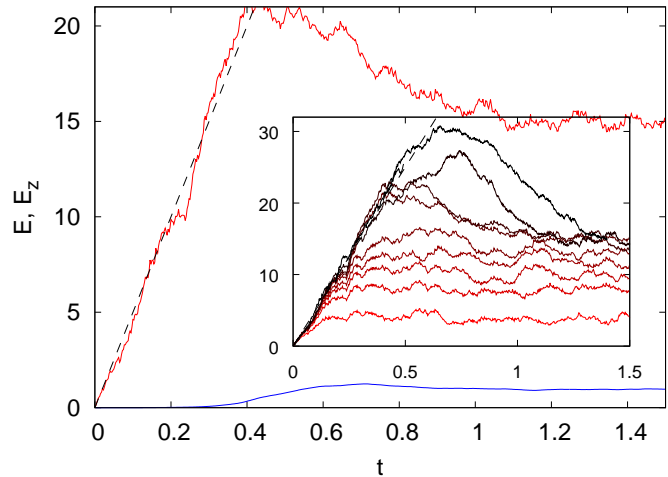}}
\caption{Time evolution of the total kinetic energy $E$ (red line) and 
of the vertical kinetic energy $E_z$ (blue line) for the forced run $6$
at $S=0.25$. The black dashed line represents the growth rate due to 
the energy input, $\varepsilon_I t$.
Inset: Time evolution of the total kinetic energy $E$ for all the runs.
Thickness $S$ (and Run number) increases from bottom (red) to top (black) 
lines. The black dashed line represents $\varepsilon_I t$.}
\label{fig1}
\end{figure}

The $2D2C$ random forcing initially produces a two-dimensional flow. 
Since it cannot transfer energy to small scale, 
energy dissipation during this first phase is negligible and the 
kinetic energy grows approximatively as $E(t) \simeq \varepsilon_I t$. 
After this phase, vertical motions start to develop and eventually produce a 
three-dimensional turbulence with a transfer of energy to small scales where 
it is dissipated. As a consequence, the kinetic energy of the flow 
reaches a (statistically) stationary state. 
This is clearly shown in Fig.~\ref{fig1} where we plot the time 
evolution of the total kinetic energy $E$ and of the vertical component
$E_z$ for the Run $6$. 
For $t \le 0.3$, the vertical kinetic energy is negligible and the 
total energy grows at the input rate. After $t \simeq 0.4$, vertical 
motion sets in and the turbulent transfer to small scales produces a 
viscous dissipation, which reduces the energy which eventually reaches 
a stationary state. We remark that in this stationary state the vertical
kinetic energy is still much smaller than the horizontal one 
(see Fig.~\ref{fig1} and Table~\ref{table1}).

From Fig.~\ref{fig1} we can clearly distinguish two phases in the 
turbulent flow: a two-dimensional regime at initial times and a 
three-dimensional regime at late times. This picture is observed 
(with quantitative differences) for all the simulations at different 
thickness as shown in the inset of Fig.~\ref{fig1}.
The fact that kinetic energy reaches a constant value indicates that,
for any thickness, the flow is unable to sustain an inverse cascade (which 
would keep the energy increasing). The reason why the {\it energy split}
scenario (in which both a direct and an inverse energy cascades are present)
is not observed in our simulations is the main result of our work and 
will be discussed in details below. 

\begin{figure}[h!]
\centerline{\includegraphics[scale=0.75]{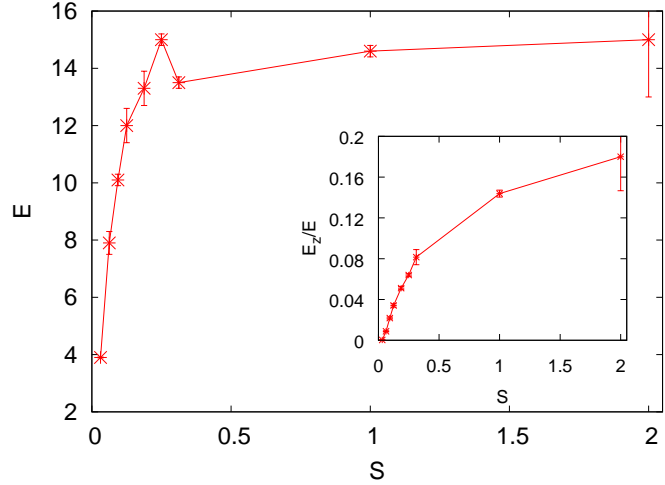}}
\caption{Asymptotic kinetic energy $E$ as a function of the thickness $S$.
Inset: ratio $E_z/E$ as a function of $S$.
}
\label{fig2}
\end{figure}

From Fig.~\ref{fig1} we observe that the asymptotic value of the energy
grows with the thickness, more rapidly for lower values of $S$. 
This is shown more clearly in Fig.~\ref{fig2}
together with the dependence of $E_z/E$ on $S$ in stationary
conditions. We see that for 
$S \lesssim 0.3$ the asymptotic energy has a strong dependence on $S$,
while for larger values of the thickness, $E$ reaches an almost constant 
plateau. 
The inset of Fig.~\ref{fig2} displays the dependence of the ratio
$E_z/E$ with the thickness. 
In this case we observe a growth for all the values of 
$S$, indicating that the presence of the bottom 
boundary affects the vertical motion even at $S \approx 1$.
We remark that the flow remains
anisotropic also at large $S$ since $E_z/E<1/3$.

Since the flow has no mean velocity, 
$\langle {\bf u} \rangle = 0$, it is natural to consider the vertical 
profiles of the rms velocities, i.e.
${\bm u}_{rms}(z)=\langle {\bm u}({\bf x})^2 \rangle^{1/2}$. The different
boundary conditions on the horizontal $(u,v)$ and vertical $w$ components,
together with the $2D$ forcing, produce different profiles for the 
horizontal and vertical components. 

\begin{figure}[h!]
\centerline{\includegraphics[scale=0.75]{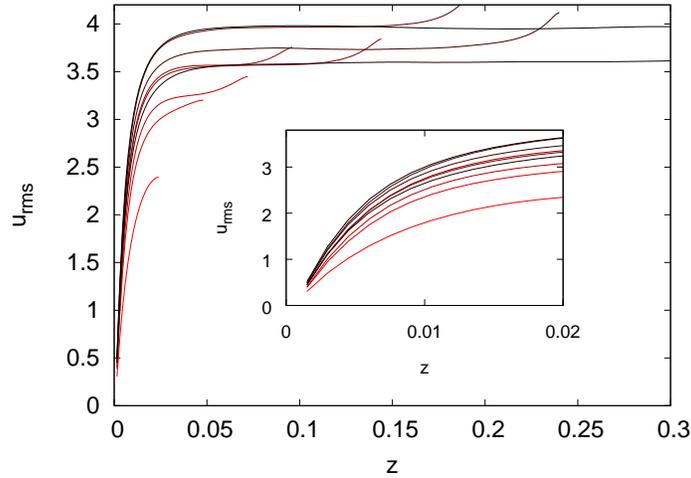}}
\caption{Vertical profile of the horizontal velocity 
fluctuations $u_{rms}$. The inset shows the behavior close to the
boundary at $z=0$. Colors as in the inset of Fig.~\ref{fig1}.}
\label{fig3}
\end{figure}

\begin{figure}[h!]
\centerline{\includegraphics[scale=0.75]{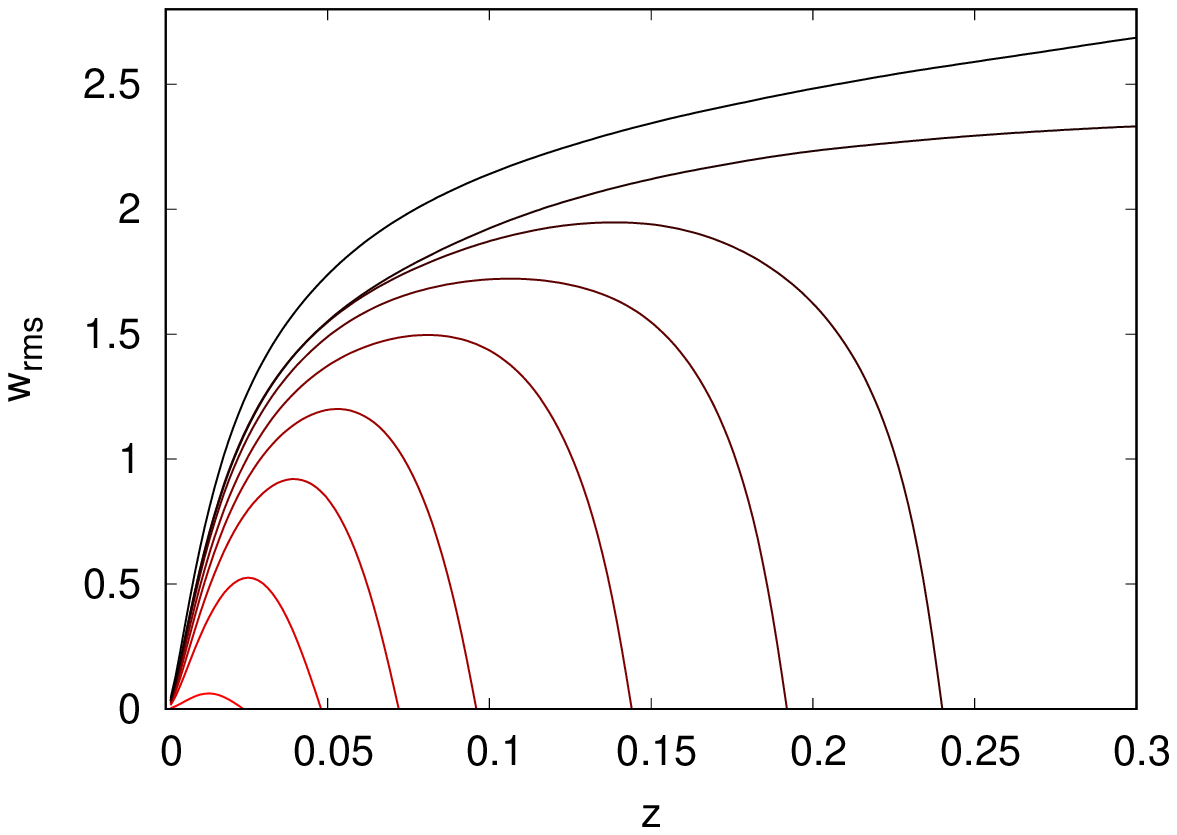}}
\caption{Vertical profile of the vertical velocity 
fluctuations $w_{rms}$. Colors as in the inset of Fig.~\ref{fig1}.}
\label{fig4}
\end{figure}

Fig.~\ref{fig3} shows the vertical profiles of one component of
the horizontal velocity $u_{rms}$ in stationary conditions. 
We observe, close to $z=0$, a boundary layer region, strongly affected
by the presence of the bottom wall, where velocity fluctuations increase
rapidly with $z$. For larger values of $z$ (and not too small
$S$) horizontal velocity fluctuations saturates to an approximately
constant value which increases with $S$. 
In these cases, we observe that close to the upper boundary $z=L_z$, 
the horizontal velocity fluctuations increase. This is a consequence 
of the free-slip boundary conditions and has been already observed in
free-surface channel flow \cite{nagaosa1999direct,lovecchio2013time}.
A possible explanation of this effect is obtained from a 
Taylor expansion of the velocity field close to the surface 
$z=L_z$. Introducing, for simplicity of notation,
 the shifted variable $Z=L_z-z$ (such that
the free surface is at $Z=0$), boundary conditions imply for 
one horizontal component of the velocity 
$u({\bf x})=u_0(x,y)+u_2(x,y)Z^2 + O(Z^4)$.
Therefore, the variance of the velocity close to the surface has the 
expression
\begin{equation}
\langle u^2 \rangle=\langle u_0^2 \rangle + 2 \langle u_0 u_2 \rangle Z^2
+ O(Z^4).
\label{eq3.1}
\end{equation}
Assuming that the energy dissipation rate averaged over 
horizontal planes is independent on $z$ (which is verified in our simulation
for values of $z$ not too close to the bottom plane), we expect that 
$\langle \varepsilon \rangle = -a \nu \langle u \nabla^2 u \rangle > 0$,
where the positive constant $a$ depends on the details of the flow 
($a=3$ for isotropic turbulence). 
By Taylor expansion we can write
\begin{equation}
u \nabla^2 u \simeq (u_0 + u_2 Z^2)(\partial_x^2+\partial_y^2+\partial_Z^2)
(u_0+u_2 Z^2) \, .
\label{eq3.2}
\end{equation}
Using integration by part in the periodic directions $x$ and $y$ we obtain
\begin{equation}
\langle u \nabla^2 u \rangle = - \langle (\partial_x u)^2 \rangle -
\langle (\partial_y u)^2 \rangle + 2 \langle u_0 u_2 \rangle + O(Z^2)
\simeq - \varepsilon/\nu < 0.
\label{eq3.3}
\end{equation}
This expression suggests that, in the absence of cancellations of leading 
terms, $\langle u_0 u_2 \rangle \propto - \varepsilon/\nu < 0$
and therefore, from (\ref{eq3.1}), that $\langle u^2 \rangle$ 
deceases moving away from the free surface as observed in 
Fig.~\ref{fig3}. 
A parabolic fit of the velocity variance close to $z=L_z$ 
is quantitatively consistent with the above predictions.
Close to the bottom boundary with no-slip BC we observe that the 
extension of the boundary layer region is weakly dependent on $S$, as it is 
shown in the inset of Fig.~\ref{fig3}.

Vertical profiles of the vertical velocity fluctuations are shown 
in Fig.~\ref{fig4}. At variance with the horizontal components, 
here the velocity (and therefore its fluctuations) vanishes also at
the upper free surface. The maximum value of fluctuations is observed
approximately in the middle of the domain, even if, due to the
different boundary conditions, the profiles are not symmetric with 
respect the central plane $z=L_z/2$.
The ratio of the lines plotted in Fig.~\ref{fig4} and in Fig.~\ref{fig3}
gives the profile of the anisotropy of the velocity field. Even for the 
largest values of $S$, we have that in the central part of the domain
$w_{rms}/u_{rms} \simeq 0.5$.

\section{Small-scale statistics and transient inverse cascade}
\label{sec4}

As discussed in the previous Section, the 2D2C forcing produces 
initially a two-dimensional flow. This flow is non-stationary, 
as shown in Figs.~\ref{fig1}, and eventually develops 
instabilities which result in a three-dimensional motion. 
In order to better understand and characterize this transition, 
we study the small scale statistics of the turbulent flow at
different times and horizontal planes. 
We consider here the run at $S=2$ for which the results are more clear, 
but a similar scenario is observed for the other thickness also. 

\begin{figure}[h!]
\centerline{\includegraphics[scale=0.75]{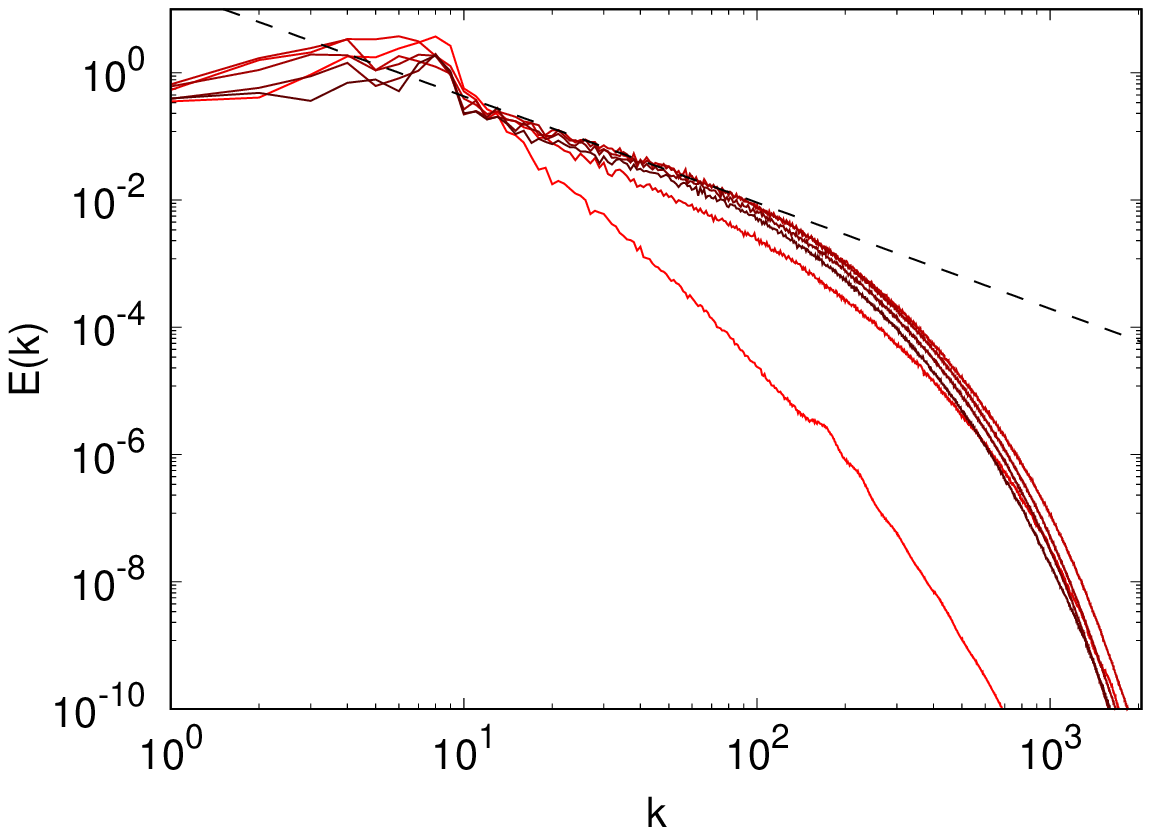}
\includegraphics[scale=0.75]{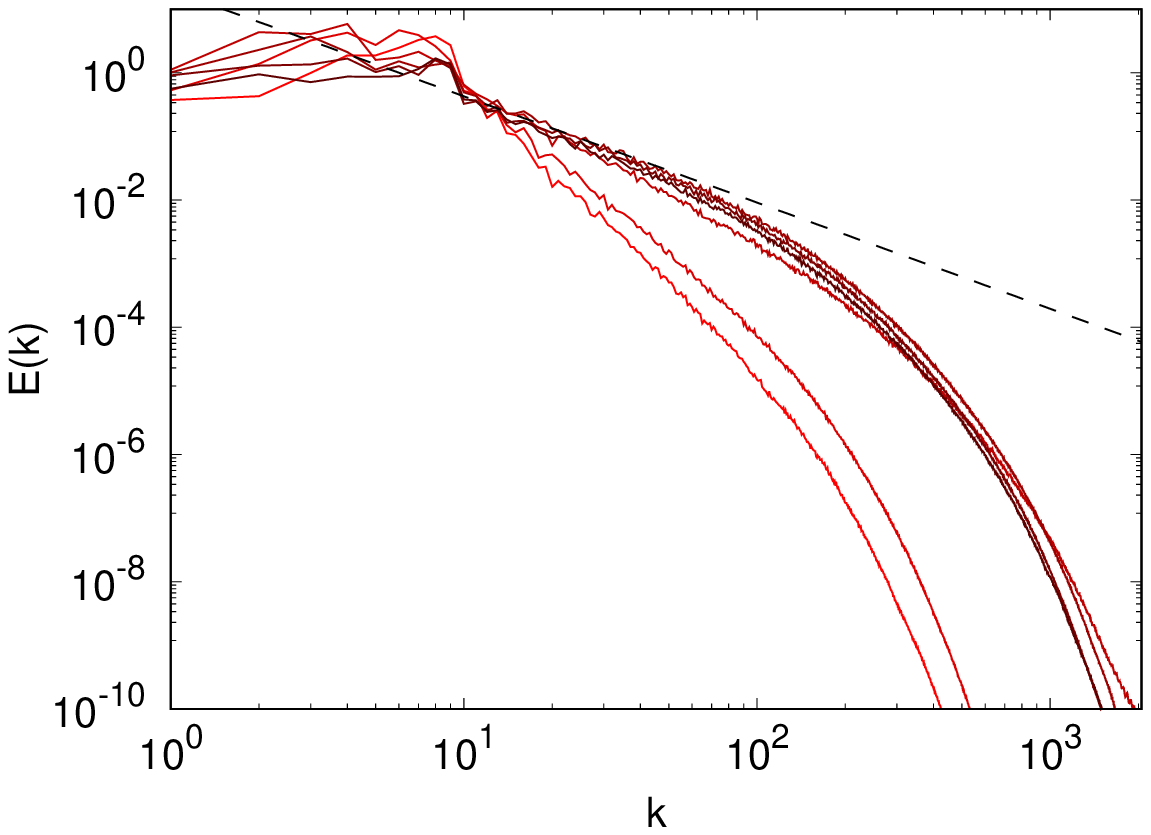}}
\caption{2D energy spectra for the simulation at $S=2$ computed at 
$z=L_z/4$ (left) and $z \simeq L_z$ (right) at different times, 
from $t=0.2$ (light red)
to time $t=1.2$ (black) with steps $\Delta t=0.2$. 
The dashed line represents the 
Kolmogorov spectrum $k^{-5/3}$.}
\label{fig5}
\end{figure}

Figure~\ref{fig5} shows the horizontal spectra (computed on the
($x,y$) plane) of the
full velocity field in two different depth $z$ and at different times. 
Both at $z=L_z/4$ and at $z \simeq L_z$ (i.e. close to the lower and 
upper boundary respectively) the flow initially (at $t=0.2$)
develops fluctuations at small scales (i.e. at $k>k_f=8$) with a 
steep spectrum compatible with a 2D direct cascade of enstrophy. 
During this first
stage, in which the energy increases approximately with the 
energy injection (see Fig.~\ref{fig1}), the flow transfers some 
energy at large scale as prescribed by a two-dimensional phenomenology. 
At $t=0.4$ the flow close to the bottom has already developed
a Kolmogorov scaling $k^{-5/3}$, compatible with a 3D direct cascade,
while the spectrum of the flow close to the surface is still steep 
(second line from bottom in both plots). 
At later times $t \ge 0.6$, spectra display a Kolmogorov scaling 
in both planes.

\begin{figure}[h!]
\centerline{\includegraphics[scale=0.75]{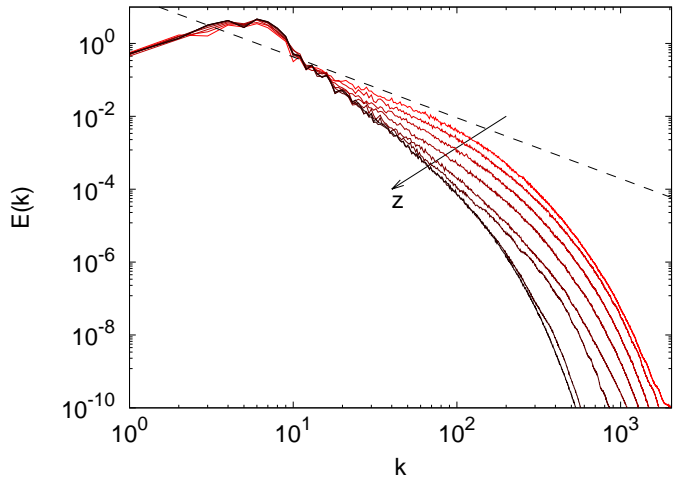}}
\caption{2D energy spectra from the simulation at $S=2$ at time $t=0.4$ 
computed at different planes from $z=L_z/8$ (clear red) to 
$L_z \simeq L_z$ (black).
The dashed line represents the Kolmogorov spectrum $k^{-5/3}$.}
\label{fig6}
\end{figure}

The interpretation of these results is that two-dimensional turbulence, 
which is initially produced by the 2D2C forcing, is transient and the 
flow develops a three-dimensional direct cascade starting from the layers
closer to the bottom boundary. In this sense, the 
transition from 2D to 3D flow is not uniform in space, but 
proceeds from the bottom layer towards the top layer. This is 
clearly shown in Fig.~\ref{fig6}, where we plot the energy spectra
computed on different horizontal planes in the domain at the 
intermediate time $t=0.4$. It is clear that, while the upper layers
(at $z \simeq L_z$) have a steep spectra compatible with a direct 
2D cascade, the lower layers close to the bottom have already 
developed a 3D cascade with a Kolmogorov spectrum.

The transition from 2D to 3D dynamics at different layers is confirmed
by the analysis of structure functions (SF) in physical space, 
in particular by the third-order SF which contains 
information about the flux of energy \cite{boffetta2012two}.

From the longitudinal velocity increments 
$\delta u_{L}(\ell,{\bm x},t) \equiv 
({\bm u}({\bm x}+{\bm \ell},t)-{\bm u}({\bm x},t)) \cdot
{\bm \ell}/\ell$
where ${\bm \ell}$ is a vector on the $(x,y)$ plane, 
we define the SF of order $p$ as
\begin{equation}
S_p(\ell;z) = \langle \delta u_{L}^p(\ell,{\bm x},t) \rangle
\label{eq4.1}
\end{equation}
where, as in Section~\ref{sec1}, the average is over the plane 
$(x,y)$ and time.
We remark that three-dimensional turbulence is characterized by a 
negative third-order SF corresponding to a direct cascade of 
turbulent fluctuations to small scales. In particular, in 
3D homogeneous-isotropic turbulence one has 
$S_3(\ell)=-(4/5) \varepsilon \ell$ 
\cite{frisch1995turbulence}.
In two dimensions one has, on the contrary, an inverse cascade of 
turbulent fluctuations at scales larger than $L_f$
with a positive third-order SF given,
in homogeneous-isotropic conditions, by
$S_3(\ell)=(3/2) \varepsilon \ell$ 
\cite{boffetta2012two}.

\begin{figure}[h!]
\centerline{\includegraphics[scale=0.75]{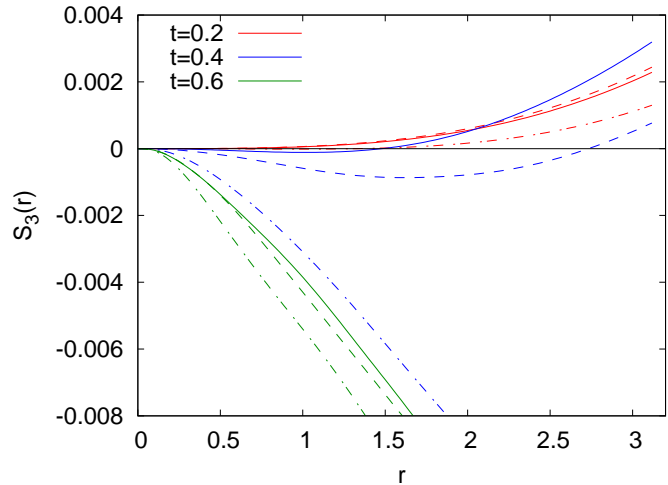}}
\caption{Horizontal longitudinal third-order velocity structure functions 
$S_3({\ell;z})$ 
computed at three times of the simulation, $t=0.2$ (red lines),
$t=0.4$ (blues lines) and $t=0.6$ (green lines). 
Continuous lines correspond to $z=3/4 L_z$, dashed line to 
$z=1/2 L_z$ and dashed-dotted line to $z=1/4 L_z$.
Simulation at $S=2$.}
\label{fig7}
\end{figure}

In Fig.~\ref{fig7} we plot the horizontal third-order longitudinal velocity
SF at three different times and at three different depth corresponding
to $z/L_z=1/4,1/2,3/4$. At short time, $t=0.2$ the SF is positive,
corresponding to an inverse energy cascade, at all the three depths considered,
consistent with the spectrum shown in Fig.~\ref{fig4}.
At the intermediate time $t=0.4$ (which still correspond to the 
growing phase of the total energy, see inset of Fig.~\ref{fig1}), the SF in the 
upper layer (continuous line) is still positive, while it becomes 
negative in the lower layer close to the bottom boundary (dotted-dashed
line). At the intemediate layer the SF change sign with the scale and
at large scales is still positive. 
This confirms the picture in Fourier space observed in Fig.~\ref{fig6}.
At time $t=0.6$, which corresponds to the peak of the kinetic
energy in Fig.~\ref{fig1}, the SFs become negative for all the 
depth considered, indicating a complete transition to 3D turbulence 
in the whole domain. 
This is at variance to what observed in simulations in fully 
periodic domains where, for intermediate values of $S$, it is observed an
inverse cascade at large scales together with a direct cascade at
small scales, both in stationary conditions \cite{celani2010turbulence}.
Remarkably, laboratory experiments in a conducting fluid 
show a similar phenomenology with the transion from positive to 
negative third-order SF \cite{gledzer2011structure}.

The physical interpretation of these results, which are qualitatively 
confirmed also for the other simulations at smaller $S$, is that 
the 2D2C forcing initially produces a quasi two-dimensional flow,
which is turbulent (positive $S_3(\ell)$) in the whole domain. 
As the energy increases, the friction on the bottom boundary 
induces vertical motions which make the flow three-dimensional
(negative $S_3(\ell)$) starting from the lower layers. 
Eventually, at longer times, the whole flow becomes three dimensional,
turbulent fluctuations are dissipated by viscosity and the kinetic
energy decreases to reach a stationary state. 

\section{Decaying turbulence}
\label{sec5}

In order to better understand the effects of the bottom layer on the 
turbulent flow, we performed additional simulations in decaying conditions,
i.e. by integrating (\ref{eq1}) without the forcing term ${\bm f}$.
The initial conditions for the decaying simulations are taken from the 
forced runs in stationary conditions, i.e. from the last time of
the inset of Fig.~\ref{fig1}. 

\begin{figure}[h!]
\centerline{\includegraphics[scale=0.75]{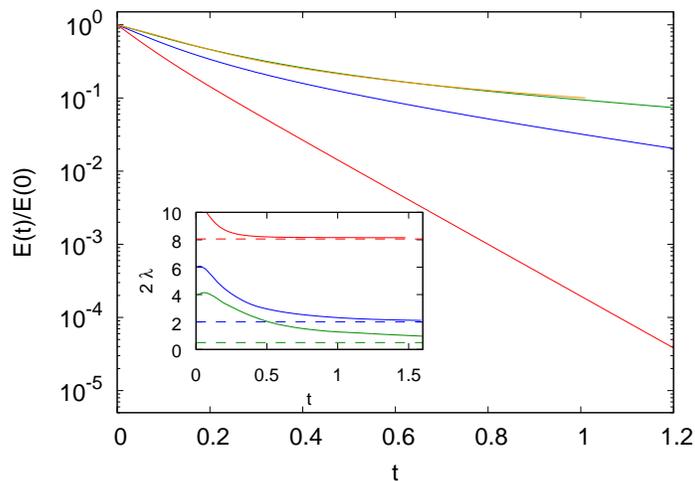}}
\caption{Decaying of total kinetic energy in unforced simulations at different 
thickness $S=0.03125$ (red lower line), $S=0.0625$ (blue intermediate line),
$S=0.125$ and $S=0.3125$ (green and orange upper lines). 
Inset: local exponential slope of $\log (E(t)/E(0))$ for simulations 
at $S=0.03125$ (red upper line), $S=0.0625$ (blue intermediate line),
$S=0.125$ (green lower line). Dashed lines represent the viscous decay
rate $\lambda=\pi^2 \nu/(4 L_z^2)$.}
\label{fig8}
\end{figure}

The evolution of the energy (normalized with the initial energy $E(0)$) is 
shown in Fig.~\ref{fig8} (where the initial time $t=0$ is now the time at 
which the forcing is switched off). From a qualitative point of view, it is
evident that the decay is faster for thinner layers, indicating the 
importance of the bottom boundary for the dissipation of energy. 
Nonetheless, we see that energy dissipation rate becomes almost independent in 
the case of thick layers (the lines for $S=0.125$ and
$S=0.3125$ are practically identical). 

The lin-log plot of Fig.~\ref{fig8} suggests that, while in the thinnest case 
$S=0.03125$ the long-time decay is with good approximation exponential, 
this is not the case for the
simulations at larger values of $S$. This is quantitatively confirmed by the 
inset of Fig.~\ref{fig8} where we plot the local rate of exponential decay for 
three cases. The rate $2 \lambda$ is obtained by plotting the local slope of 
$\log (E(t)/E(0))$ vs time, while the dashed lines represent the 
theoretical viscous decaying rate of a 2D flow with friction, given by
$\lambda=\nu \pi^2/(4 L_z^2)$ 
\cite{shats2010turbulence}.
It is clear that, while the flow at $S=0.03125$, after a short transient, 
reaches the exponential decay with the predicted viscous rate, the other cases 
with larger $S$ display a more complex decay law which cannot be 
simply described by an exponential law and a corresponding  friction 
coefficient $\lambda$. A complex decaying law in quasi 2D experiments
has been recently reported and interpreted as different stages of 
exponential decay with different decaying constant \cite{fang2017multiple}.
The results shown in the inset of Fig.~\ref{fig8} indicate that in our case
it is difficult to recognize a, even transient, exponential regime and 
that the interplay of 2D and 3D motion produces a complex decaying 
phenomenology.

\section{Conclusions}
\label{sec6}

We studied the dynamics and statistics of turbulence in a thin layer
with no-slip BC on the bottom surface, forced by a 2D2C forcing. 
For the range of thickness explored, we find that the flow is 
unable to sustain an inverse energy cascade: after a initial transient, 
the flow develops a three-dimensional direct cascade which starts from the 
bottom and eventually propagates in the whole domain.
The analysis of the energy spectrum and of the third-order structure 
function shows that at intermediate times, a 2D like and a 3D like
phenomenologies can coexist at different depth in the flow.
This is in contrast to what observed in homogeneous simulations in 
the absence of boundaries where the 
the flow can sustain simultaneously an inverse cascade of energy to 
large scales and a direct cascade to small scales \cite{celani2010turbulence}.
Moreover, our results are also at odds with several laboratory
experiments where an inverse cascade is observed 
\cite{paret1997experimental,chen2006physical,von2011double,byrne2011robust,xia2011upscale}, 
but mostly in the presence of two layers of fluids (miscible
or not). The differences in these cases are probably due to the fact
that the dynamics in the upper layer (where the flow is studied), is 
partially decoupled from the lower layer which 
is affected by the no-slip boundary conditions.

Finally, we studied the decaying behavior of the thin turbulent layer. 
Also in this case we find differences with respect to laboratory 
experiments where the decay of kinetic energy is exponential and 
therefore it is parameterized by a single friction coefficient 
\cite{shats2010turbulence}. 
We observe a clear exponential decay only
for the simulation with the thinnest layer, while in the other cases 
a more complex decay law is observed. 
The different behavior in this case can be not only ascribed to the 
presence of stratification, but also to the 
different initial flow conditions between experiments and simulations. 

In this work we have changed only one parameter of the flow (the 
thickness $S$), while there are other variables which can produce
a different phenomenology, mainly the viscosity (i.e. the Reynolds
number) and the forcing statistics \cite{poujol2020role}. 
Further work is therefore needed to 
uncover the reach phenomenology of turbulent thin layers.

\section*{Acknowledgments}
We acknowledge HPC CINECA for computing resources
(INFN-CINECA grant no. INFN22-FieldTurb).
M.E.R. is supported by the Okinawa Institute of Science and Technology Graduate University (OIST) with subsidy funding from the Cabinet Office, Government of Japan. M.E.R. also acknowledges the computational time provided by the Scientific Computing section of Research Support Division at OIST.

\bibliography{biblio}

\end{document}